\documentclass[twocolumn,showpacs,preprintnumbers]{revtex4}
\usepackage{amssymb}
\usepackage{amsmath}
\usepackage{graphicx}
\usepackage{dcolumn}
\usepackage{bm}
\usepackage{color}

\begin{document}

\title{Implication of giant photon bunching on quantum phase transition in the dissipative anisotropic quantum Rabi model}
\author{Tian Ye$^{1,2}$}
\author{Chen Wang$^{3,}$}
\email{wangchen@zjnu.cn}
\author{Qing-Hu Chen$^{1,2,}$}
\email{qhchen@zju.edu.cn}

\address{
$^{1}$ Zhejiang Province Key Laboratory of Quantum Technology and Device, School of Physics, Zhejiang University, Hangzhou 310027, China \\
$^{2}$  Collaborative Innovation Center of Advanced Microstructures, Nanjing University, Nanjing 210093, China\\
$^{3}$ Department of Physics, Zhejiang Normal University, Jinhua 321004, Zhejiang, China
 }\date{\today }

\begin{abstract}
We investigate  the quantum phase transition in the dissipative anisotropic quantum Rabi model in the framework of quantum dressed master
equation.
From perspectives of both numerical and  analytical analysis, we  unravel the implication of the giant photon-bunching feature on the first-order quantum phase transition.
The observed two-photon statistics can be well described analytically within a few lowest eigenstates at the low temperature.
Moreover, such significant photon-bunching peak is generally exhibited at the deep-strong qubit-photon coupling, which is however lacking in the dissipative isotropic quantum Rabi model.
Therefore, we suggest that the photon-bunching measurement is helpful to characterize the first-order QPT of the qubit-photon hybrid systems.
\end{abstract}

\pacs{42.50.Ct, 42.50.Ar, 03.65.Yz}
\maketitle

\section{\label{Introduction} Introduction \newline
}

\indent Deep understanding and smart manipulation of quantum light-matter
interaction becomes a challenging interdisciplinary frontier, which is
scientifically important and practically demanding in the quantum community~%
\cite{scully1997book,haroche2006book}. One representative paradigm to
theoretically describe quantum light-matter interaction is the qubit-photon
coupled quantum system~\cite{wallquist2009pst,kurizki2015pnas}, i.e., a
multitasking platform for complementary functionalities with different
physical components (e.g., qubit and optical resonator), which has been
extensively investigated in diverse fields, ranging from quantum optics~\cite%
{diaz2019rmp,kockum2019nrp}, quantum information~\cite%
{blais2020np,clerk2020np}, quantum phase transition (QPT)~\cite%
{hwang2015,Lin2017,xychen2021pra,duan2021nc,Ying2021}, to quantum transport~%
\cite{rozani2018np,senior2020cp,pekola2021rmp,cwang2021pra}. The
experimental realizations of the ultrastrong qubit-photon coupling~\cite%
{niemczyk2010np,yoshihara2017np,anappara2009prb} stimulate the researchers
to explore uncharted regimes of qubit-photon quantum systems.

The quantum Rabi model (QRM)~\cite%
{rabi_rabimodel,braak2011prl,chen2012pra,Zhong13,braak2016jpa} is widely
considered as the generic model to characterize qubit-photon  systems,
which is composed of one qubit interacting with a single-mode radiation
field. Traditionally, the qubit-photon coupling in the quantum
electrodynamics platforms, e.g., natural atoms in resonant cavity, is rather
weak~\cite{haroche2020np}, which reduces the QRM to the Jaynes-Cummings
model \cite{jcm1963ieee} (JCM) by performing the
rotating-wave approximation.
However, along with the dramatic advances  of solid-state quantum platforms%
, such as circuit quantum electrodynamics systems ~\cite%
{ov2011prl,baust2016prb,yoshihara2017np,stockklauser2017prx,bayer2017nanolett}, the
intriguing ultrastrong interaction has attracted tremendous attention ~\cite%
{diaz2019rmp,kockum2019nrp,blais2020np}, where the qubit-photon coupling
strength is comparable to the individual components energies. This directly
invalidates the rotating-wave approximation. Consequently, the
counter-rotating-wave (CRW) terms produce a series of exciting physical
phenomena, e.g., parity symmetry restricted photon-wave-packet propagation~%
\cite{jc2010prl}, vacuum Rabi splitting \cite{yyzhang2013cpl}, one photon
simultaneously exciting two atoms \cite{lg2016prl}, and multiphoton
sidebands transition \cite{jqyou2017pra}.

Aimed to bridge the gap between the JCM and QRM, the anisotropic quantum
Rabi model (AQRM) was  proposed \cite{Ye13,qtxie2014prx}, where the
strengths of the rotating-wave (RW) and CRW terms are relaxed to be
independent. In the weak CRW terms limit, the AQRM is shifted to the JCM,
whereas the AQRM becomes QRM with identical coupling strengths for both RW
and CRW terms. The AQRM nowadays can be realized by various quantum
platforms, such as inductively coupled circuit quantum electrodynamics
systems \cite{ab2014prl,wjy2017pra,gcwang2019scirep}, two-dimensional
electron gas with spin-orbit coupling \cite{erlingsson2010pra}, and a
two-level qubit exchanging spin with one anisotropic ferromagnet \cite%
{skogvoll2021arxiv}.
{\ By exploring the energy spectrum of the AQRM,} the first-order QPT is
observed when the CRW terms are weaker than RW terms~\cite{qtxie2014prx},
which is, in sharp contrast, absent in the QRM. Very recently, Chen \emph{et
al}. found  that the AQRM undergoes infinitely many first-order QPT by
continuously increasing the qubit-photon coupling strength~\cite%
{xychen2021pra}.

In reality, the qubit-photon  quantum system inevitably interacts with
the environment~\cite{uweiss2012book}. The system-bath interaction induced
quantum dissipation will dramatically renew both transient and steady-state
behaviors of  quantum systems, e.g., JCM and QRM, which results in
dissipative QPT~\cite%
{schmidt2010prb,hwang2016prl,hwang2018pra,hjc2015prx,jfink2017prx}. While
from the photon statistics perspective, quantum dissipation enables the
reasonable measurements of the nonclassicality~\cite{hjc1999book}%
. Two-photon correlation function is widely considered as the main
observable to characterize the nonclassicality of output photons, which is
originally introduced by  Glauber to investigate the optical coherence of
quantum theory~\cite{rjg1963pr}.
{However, due to the realizations of
ultrastrong qubit-photon coupling~\cite{diaz2019rmp,kockum2019nrp}, the
system-bath interaction should be properly treated in the eigenstate
framework of the whole qubit-photon quantum systems~\cite%
{blais_dME,alb2017pra,as2018pra,alb2020aqt}. In particular, Ridolfo \emph{et
al}. recently proposed a modified definition of the two-photon correlation
function to characterize steady-state photon statistics in the dissipative
QRM~\cite{ridolfo_photonblockade,ar2013prl,alb2016pra,lg2017acsp}, based on the quantum
dressed master equation (DME)~\cite{blais_dME}.
Therefore, we naturally raise the question: Could the nonclassicality of photons of the AQRM
coupled to the thermal baths at low temperature keeps the trace of the first-order QPT in
the closed AQRM?

In this work, we apply the modified expression of the two-photon correlation
function in Ref.~\cite{ridolfo_photonblockade} to investigate quantum phase transition in the dissipative AQRM,
where the dissipative dynamics is characterized by the DME. We will focus on the effect of the  anisotropic
qubit-photon interaction on  two-photon statistics in the
deep-strong coupling regime. We also reveal the implication of the multiple first-order QPTs with the observed sharp photon bunching behavior in the analytical sense.
The paper is
organized as follows: In Sec. II, we give a brief introduction to the
anisotropic quantum Rabi model, the dressed master equation and two-photon
correlation function. In Sec. III, we  analyze the emerging sharp photon-bunching feature
and the implication on  the first-order quantum phase transition. We give
a conclusion in the last section.}
\section{\label{mm}Model and Method}

\subsection{\label{model} Anisotropic quantum Rabi model}

{\ The Hamiltonian of AQRM is described as} \cite{qtxie2014prx,xychen2021pra}
\begin{eqnarray}
H _\text{AQRM}&=&\omega _{0}a ^{\dagger}a +\frac{\Delta}{2}\sigma _{z}
\notag \\
&&+g\left[(a\sigma _{+} +a^{\dagger}\sigma _{-}) +r(a\sigma _{-}
+a^{\dagger}\sigma _{+})\right],  \label{AQRM}
\end{eqnarray}%
where $a^{\dagger}$ ($a$) creates (annihilates) a photon in the cavity with
the frequency $\omega_0$, $\Delta $ is the energy splitting of qubit, $%
\sigma_{\pm} =\frac{1}{2}(\sigma_{x}\pm i\sigma_{y})$ excites (relaxes) the
qubit from the state $|0(1){\rangle}$ to $|1(0){\rangle}$, with $%
\sigma_{\alpha=x, y, z}$ the Pauli operators and $|0(1){\rangle}$ the qubit
ground (excited) state, $g$ is the qubit-photon coupling strength, and $r$
is the  anisotropic parameter.

{When $r=1$, the AQRM is reduced to QRM. Furthermore, as the qubit-cavity
coupling becomes weak, the CRW terms $(a\sigma _{-}+a^{\dagger }\sigma _{+})$
become negligible. Under the rotating-wave approximation, the QRM is reduced
to the JCM \cite{jcm1963ieee}, which is identical with the $r=0$ limit.
Here, the AQRM ties the QRM and the JCM to investigate the nontrivial
influence of CRW terms.} Moreover, the AQRM possesses the same symmetry as
the QRM, i.e., $\mathbb{Z}_{2}$ symmetry. Specifically, the eigenstate $%
|\phi _{n}\rangle $ of the system Hamiltonian $H_{\text{AQRM}}$ satisfies $%
-\sigma _{z}e^{i\pi a^{\dagger }a}|\phi _{n}\rangle =\Pi |\phi _{n}\rangle $%
, with the parity eigenvalue $\Pi =\pm 1$. Hence, we are able to classify
the eigenstates in the even and odd parity subspaces, which correspond to $%
\Pi =1$ and $\Pi =-1$, respectively.\newline

{\indent
Due to the tremendous advances of the quantum technology, the AQRM with ultrastrong qubit-photon coupling can be realized in superconducting circuits,
e.g., based on two inductively interacting superconducting quantum interference devices (SQUIDs)~\cite{qtxie2014prx}.
Specifically, a primary SQUID with the comparative large loop generates an electromagnetic field,
which is inductively coupled to the second SQUID qubit.
In the limit of ignored capacitive interaction between these two SQUIDs,
the anisotropic qubit-photon coupling can be steadily realized and operated by including both inductance of the circuit and the mutual inductance.
}

\subsection{\label{ME} Quantum dressed master equation}

Practically, the  quantum system inevitably interacts with the
environment. To describe the influence of thermal baths on the AQRM, i.e.,
the qubit and the photon field individually coupled to two bosonic thermal
baths, the total Hamiltonian is expressed as
\begin{equation}
H_{\text{total}}\ =H_{\text{AQRM}}+H_{B}+V,
\end{equation}%
where the first term is just the AQRM Hamiltonian (\ref{AQRM}). Two bosonic
thermal baths are described as $H_{B}=\Sigma _{u=q,c;k}\omega
_{k}b_{u,k}^{\dagger }b_{u,k}$, where $b_{u,k}^{\dagger }$ ($b_{u,k}$) is
the annihilation (creation) operator of bosons with the frequency $\omega
_{k}$ in the $u$th bath. The interaction between the AQRM and two bosonic
thermal baths is denoted as $V=V_{q}+V_{c}$, where two interaction terms
associated with the qubit and the photon field read
\begin{subequations}
\begin{align}
V_{q}=& \Sigma _{k}\lambda _{q,k}(b_{q,k}+b_{q,k}^{\dagger })\sigma _{x},~
\label{vq} \\
V_{c}=& \Sigma _{k}\lambda _{c,k}(b_{c,k}+b_{c,k}^{\dagger })(a+a^{\dagger
}),~  \label{vc}
\end{align}%
with $\lambda _{q,k}~(\lambda _{c,k})$ the coupling strength between the
qubit (photon field) and the corresponding bosonic thermal bath. Generally,
the system-bath interaction can be characterized as the spectral function $%
\gamma _{q(c)}(\omega )=2\pi \sum_{k}|\lambda _{q(c),k}|^{2}\delta (\omega
-\omega _{k})$. In this work, we apply the Ohmic spectrum case to quantify
the system-bath interaction~\cite{uweiss2012book}, i.e., $\gamma _{q}(\omega
)=\alpha _{q}{\omega }\exp (-{\omega }/{\omega _{c}})/\Delta $ and $\gamma
_{c}(\omega )=\alpha _{c}{\omega }\exp (-{\omega }/{\omega _{c}})/{\omega
_{0}}$, where $\alpha _{q(c)}$ is the system-bath coupling strength, and $%
\omega _{c}$ is the cutoff frequency of bosonic thermal bath.

{Under the assumption that the system-bath interaction is weak, we can
separately perturb both $V_c$ and $V_q$ to obtain the DME by including
Born-Markov approximation, which may be proper to handle long-time dissipative dynamics at ultra-strong and deep-strong qubit-photon couplings~\cite{blais_dME,alb2020aqt}.
Indeed, as the interaction between the qubit and photons goes beyond strong,
the light-matter system should be treated as a whole~\cite{diaz2019rmp,kockum2019nrp}.
Accordingly, the system-bath interactions (\ref{vq})-(\ref{vc}) should be described under the eigenmodes of the AQRM,
i.e., $V_{q}=\sum _{k,m,n}\lambda _{q,k}(b_{q,k}+b_{q,k}^{\dagger })P^q_{nm}$
and $V_{c}=\sum _{k,m,n}\lambda _{c,k}(b_{c,k}+b_{c,k}^{\dagger })P^c_{nm}$,
with the eigenmode projectors $P^q_{nm}={\langle}\phi_n|\sigma_x|\phi_m{\rangle}|\phi_{n}\rangle\langle\phi_{m}|$
and $P^c_{nm}={\langle}\phi_n|(a^\dag+a)|\phi_m{\rangle}|\phi_{n}\rangle\langle\phi_{m}|$,
and the eigenstate of the AQRM $|\phi_{m(n)}\rangle$.
This ensures that zero-temperature thermal baths will drive the quantum system to the ground state.
}
Finally, the DME is described as
\end{subequations}
\begin{eqnarray}~\label{dme1}
~ \frac{\partial\rho_{s}}{\partial t}&=&-i[H_{\text{AQRM}},\rho_{s}]  \notag
\label{dme} \\
&&+\sum_{j,k>j} \{ \Gamma_u^{j,k} \left[1+n_u\left(\Delta_{k,j}\right)\right]
\mathcal{D}[|\phi_j\rangle\langle\phi_k|,\rho_{s}]  \notag \\
&&+\Gamma_u^{j,k}n_u\left(\Delta_{k,j}\right) \mathcal{D}[|\phi_k\rangle%
\langle\phi_j|, \rho_{s}] \},
\end{eqnarray}
where the dissipator is $\mathcal{D}[O,\rho_{s}] =\frac{1}{2}(2 O\rho_{s}
O^\dagger -O^\dagger O\rho_{s} -\rho_{s}O^\dagger O)$, $\rho_{s}$ is the
reduced density matrix of the AQRM, $\Delta _{k,j}=E_{k}-E_{j}$ is the
energy gap between two eigenstates (i.e, $|\phi_k{\rangle}$ and $\phi_j{%
\rangle}$), $n_u(\Delta_{k,j})={1}/[{\exp(\Delta_{k,j}/k_B T_u)-1}]$ is the
Bose-Einstein distribution function, with $k_B$ the Boltzmann constant and $%
T_u$ the temperature of the $u$-th thermal bath, and the transition rates $%
\Gamma^{k,j}_{q}$ and $\Gamma^{k,j}_{c}$ are given by
\begin{subequations}
\begin{align}
\Gamma_q^{k,j}=&\alpha_q \frac{\Delta_{k,j}}{\Delta} e^{-\frac{\Delta_{k,j}}{%
\omega_{c}}}|\langle\phi_j|(\sigma_- +\sigma_+)|\phi_k\rangle|^2, \\
\Gamma_c^{k,j}=&\alpha_c\frac{\Delta_{k,j}}{\omega_0}e^{-\frac{\Delta_{k,j}}{%
\omega_{c}}}|\langle\phi_j|(a+a^\dagger)|\phi_k\rangle|^2.
\end{align}
 Note that  only $\sigma_{\pm}$ and linear photon operator  appear  in
these expressions, the transitions are constrained between the eigenstates
with different parities.
{Based on Eq.~(\ref{dme1}), the steady state of AQRM becomes the ground state under the condition of $T_q=T_c=0$.
}

{In this work, we perform the numerical diagonalization in each parity subspace of the AQRM to obtain the   eigenstates  and eigen-energies needed in the DME. One can note that the very high excited states are actually not involved.  Even in the deep-strong coupling regime considered below,  the convergence for both the eigenstates  and eigen-energies employed in the calculation  can be achieved very well within the truncated photon number around two hundred in the Fock-state basis.}

\subsection{\label{CF}Two-photon correlation function}

In quantum optics, the two-photon correlation function is widely considered as
one powerful utility to measure the nonclassical radiation of photons, which
also reflects the intrinsic correlation properties of quantum materials in
light-matter interacting systems~\cite%
{scully1997book,qs2019prl,shc2020np,asp2020np,qbin2020prl}. The two-photon
correlation function  is defined as
\cite{hjc1999book,rjg1963pr}
\end{subequations}
\begin{equation}
G_{2}(\tau )=\frac{\langle a^{\dagger }(t)a^{\dagger }(t+\tau )a(t+\tau
)a(t)\rangle }{\langle a^{\dagger }(t)a(t)\rangle ^{2}},\label{g2_old}
\end{equation}%
where $\langle \cdots \rangle $ stands for the mean value under the $t$-time
reduced system density matrix. When the qubit-cavity coupling of the AQRM is
sufficient weak, the output field is approximately proportion to the
intra-cavity field. Thus, the photon statistic of the output field at the steady
state can be properly described by Eq.~(\ref{g2_old}).
However, such definition may break down in the ultrastrong qubit-photon coupling regime~%
\cite{ridolfo_photonblockade}.

To characterize the two-photon statistics at strong qubit-photon coupling,
{ Ridolfo \emph{et al.} derived a modified input-output relation in the dressed-state basis,
i.e., $A_{out}=A_{in}-i\sqrt{\kappa}X^{+}$~\cite{ridolfo_photonblockade},
where $A_{in(out)}$ denotes the input (output) radiation field operator, $\kappa$ is the loss rate
 of the photons via the interaction with the external detection modes,
 and the detection operator $X^{+}$ in the dressed-state basis is
\begin{equation}
{X}^{+}=-i\sum_{j,k>j}\Delta _{k,j}X_{j,k}|\phi _{j}\rangle \langle \phi
_{k}|,
\end{equation}%
with $X_{j,k}=\langle \phi _{j}|(a+a^{\dagger })|\phi _{k}\rangle$ and $%
X^{-}=(X^{+})^{\dagger }$.
This treatment naturally overcomes the inconsistency that finite photon current unphysically occurs at the ground state,
i.e., $X^{+}|\phi_0{\rangle}=0$.
By setting the  input field as vacuum,
the output photon flux emitted by the optical resonator is expressed as
$I_{out}=\kappa{\langle}X^-(t)X^+(t){\rangle}$.
Similarly, the  output
delayed coincidence rate is proportional to the two-photon
correlation term ${\langle}X^-(t)X^-(t+\tau)X^+(t+\tau)X^+(t){\rangle}$.
Hence,
the two-photon correlation function is
redefined as
\begin{equation}
G_{2}(\tau )=\frac{\langle X^{-}(t)X^{-}(t+\tau )X^{+}(t+\tau
)X^{+}(t)\rangle }{\langle X^{-}(t)X^{+}(t)\rangle ^{2}},
\end{equation}%
where ${\langle \mathcal{A}\rangle }=\text{Tr}\{\rho _{s}(t)\mathcal{A}\}$
stands for  the mean value at the reduced density matrix $\rho _{s}(t)$.
Consequently, the two-photon correlation function has been applied broadly to characterize the two-photon statistics
~\cite{ridolfo_photonblockade,ar2013prl,alb2016pra,lg2017acsp}.}
While for the representative zero-time delay case
at the steady state, $G_{2}(\tau )$ is reduced to
\begin{equation}
~G_{2}(0)=\frac{\langle X^{-}X^{-}X^{+}X^{+}\rangle _{ss}}{\langle
X^{-}X^{+}\rangle _{ss}^{2}}, \label{g20}
\end{equation}%
where $\langle \cdots \rangle _{ss}$ denotes the expectation value at the steady
state. {It should be emphasized that the influence of the two-photon correlation measurement on the steady state is negligible due to much weaker coupling assumption compared to the system-environment interactions.} As is well known, the bunching and antibunching features of photons
are two representative non-classical phenomena of light \cite%
{scully1997book,rjg1963pr,ridolfo_photonblockade}. The bunching behavior is
characterized as $G_{2}(0)>1$, which shows the super-Poisson distribution. $%
G_{2}(0)<1$ denotes antibunching behavior of photons, corresponding to the
sub-Poisson distribution. {\ $G_{2}(0)=2$ in the photon thermal field,
whereas $G_{2}(0)=1$ in the photon coherent state.}

\section{Results and discussions}

\begin{figure}[tbp]
\centering	\includegraphics[width=0.48\textwidth]{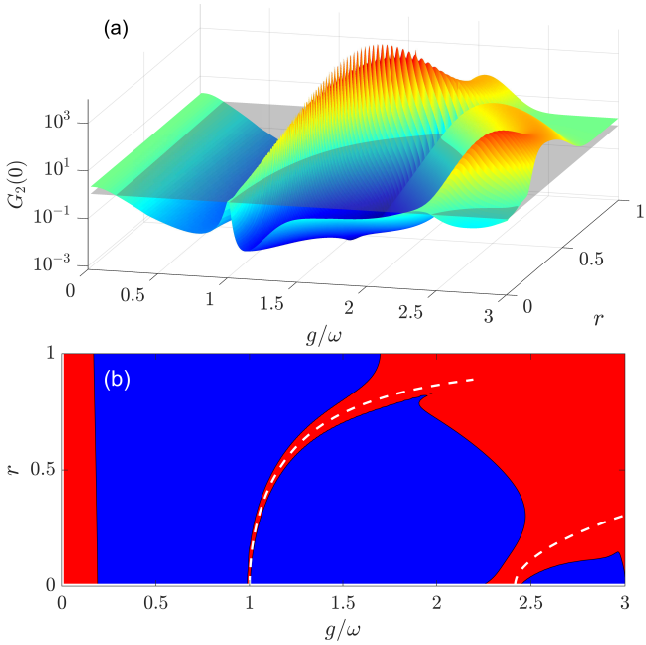}
\caption{(color online) (a) Three-dimensional view for the two-photon correlation function $G_{2}(0)$ of
the dissipative AQRM at resonance $\protect\omega _{0}=\Delta $. For
comparison, we show the gray plane of $G_{2}(0)=1$ to distinguish bunching
and antibunching of photons. (b)  Contour of $G_2(0)$ with antibunching and bunching marked in
blue and red, respectively. The white dashed
lines indicate the first-order QPTs of the closed AQRM.
The other parameters are
given by $\protect\alpha _{q}=\protect\alpha _{c}=10^{-4}\protect\omega _{0}$%
, $\protect\omega _{c}=10\protect\omega _{0}$, and $k_{B}T_{q}=k_{B}T_{c}=0.07\protect%
\omega _{0}$.}
\label{fig1}
\end{figure}

\begin{figure}[tbp]
\centering	\includegraphics[width=0.5\textwidth]{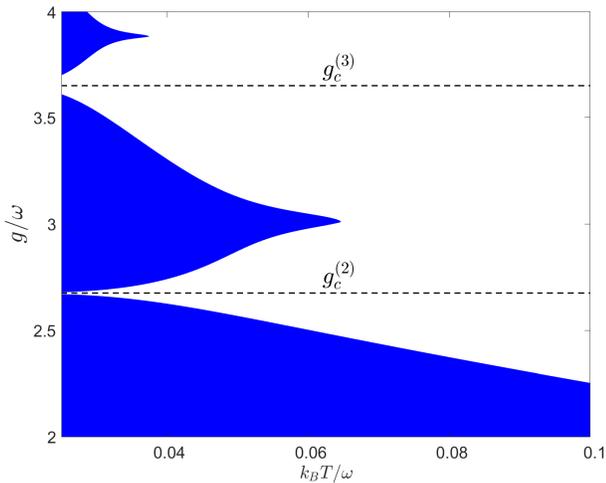}
\caption{(color online) The rescaled scheme of $G_2(0)$ with varying
temperatures of thermal baths $T_{q} =T_{c} =T$ for AQRM at $r=0.2$. Here
photon antibunching (bunching) is denoted by the blue (white) color.
The black dashed horizontal lines indicate  $g_c^{(n)}$ of the  first-order QPTs in the closed AQRM. The other system parameters are the same as in Fig.~%
\protect\ref{fig1}.}
\label{fig4}
\end{figure}

Xie \emph{et al.}  observed the first level crossing  in the AQRM  at $g_{c}^{(1)}=\sqrt{{\omega _{0}\Delta }/({1-r^{2}})}$
using the Bargmann representation ~\cite{qtxie2014prx} . At the level crossing, the parity of the ground state changes
sign and the system undergoes a first-order  QPT.  Within the Bogoliubov operator approach,
Chen \emph{et al.}~\ \cite{xychen2021pra} further revealed that the ground
state and the first excited state can cross several times, indicating
multiple first-order QPTs at $~g_{c}^{(n)}$ with $n$ a positive integer.
Recently, Fink \emph{et al.}\ experimentally characterized the first-order QPT by
measuring two-photon correlation function in the driven-dissipative
Bose-Hubbard system \cite{afink2018np}. Then, it is interesting to see
whether the first-order QPTs in the AQRM under the quantum dissipation can
also be signified by the two-photon correlation function.

We first investigate the influence of the anisotropic qubit-photon  coupling
on the two-photon correlation function $G_{2}(0)$ at the steady state in
terms of Eq.~(\ref{g20}) at low temperature ($k_{B}T_{c}=k_{B}T_{q}=0.07\omega _{0}$).
In Fig.~\ref{fig1} (a), we plot $G_{2}(0)$ as a function of $g/\omega _{0}$ and $%
r$ in a three-dimensional view.
To be clearer, we also present the  contour of $G_2(0)$ with antibunching and bunching area in Fig.~\ref{fig1} (b). Surprisingly, there exists a significant photon-bunching peak in the middle of Fig.~\ref%
{fig1} (a) within the anisotropic parameter range $0<r{\lesssim }0.8$ . In particular, the behavior of
the first critical point $g_{c}^{(1)}$~ by tuning the anisotropic parameter $r$
agrees well with the narrow photon-bunching regime, as indicated by the
white dashed line in Fig.~\ref{fig1} (b). Moreover, the divergent trend of the critical point $g_c^{(1)}$ when the anisotropic parameter $r$ is approaching unit showed by Chen et al.~\cite{xychen2021pra} is also matched with the vanishing of narrow photon-bunching peak near isotropic situation.

Next, we derive  an approximate expression of the two-photon correlation
function at the steady state analytically. In the low temperature regime, the general
relation of steady-state populations becomes $P_{0}{\gg }P_{1}{\gg }P_{2}{%
\gg }P_{3}{\cdots }$ [$P_{n}{\propto }\exp {(E_{n}/k_{B}T)}$], under the
assumption of finite energy-level spacing $\Delta _{kj}{\gg }k_{B}T$ [$%
\Delta _{kj}=(E_{k}-E_{j})$, with $E_{k}>E_{j}$ and ${H}_{\text{AQRM}}|\phi
_{k}{\rangle }=E_{k}|\phi _{k}{\rangle }$]. For simplicity, the Hilbert
space of AQRM is truncated to the subspace spanned by four lowest
eigenstates of the AQRM, i.e., the ground state $|\phi _{0}\rangle $, and
three low exited states $|\phi _{1}\rangle $, $|\phi _{2}\rangle $, and $%
|\phi _{3}\rangle $. Accordingly, the one-photon correlation term is
estimated as $\langle X^{-}X^{+}\rangle _{ss}{\approx }\Delta
_{1,0}^{2}|X_{0,1}|^{2}P_{1}$, which is contributed by the main transition $%
|\phi _{0}{\rangle }{\rightarrow }|\phi _{1}{\rangle }$. And the two-photon
correlation term is given by $\langle X^{-}X^{-}X^{+}X^{+}\rangle _{ss}{%
\approx }\sum_{j<k<l,l{\leq }3}\Delta _{kj}^{2}\Delta
_{lk}^{2}|X_{jk}X_{kl}|^{2}P_{l}$, which is generally composed by four
cooperative transitions, i.e., $|\phi _{0}{\rangle }{\rightarrow }|\phi _{1}{%
\rangle }{\rightarrow }|\phi _{2}{\rangle }$, $|\phi _{0}{\rangle }{%
\rightarrow }|\phi _{1}{\rangle }{\rightarrow }|\phi _{3}{\rangle }$, $|\phi
_{0}{\rangle }{\rightarrow }|\phi _{2}{\rangle }{\rightarrow }|\phi _{3}{%
\rangle }$, and $|\phi _{1}{\rangle }{\rightarrow }|\phi _{2}{\rangle }{%
\rightarrow }|\phi _{3}{\rangle }$. Consequently, the two-photon correlation
function is approximated as
\begin{eqnarray}
~G_{2}(0) &\approx &\frac{1}{\Delta _{1,0}^{4}|X_{0,1}|^{4}P_{1}^{2}}{\times
}\Big\{\Delta _{1,0}^{2}|X_{0,1}|^{2}\Delta _{2,1}^{2}|X_{1,2}|^{2}P_{2}
\notag \\
&&+[(\Delta _{2,0}^{2}|X_{0,2}|^{2}+\Delta _{2,1}^{2}|X_{1,2}|^{2})\Delta
_{3,2}^{2}|X_{2,3}|^{2}  \notag \\
&&+\Delta _{1,0}^{2}|X_{0,1}|^{2}\Delta _{3,1}^{2}|X_{1,3}|^{2}]P_{3}\Big\},
\label{g2approx}
\end{eqnarray}%
which tightly relies on the transition coefficient $X_{kj}$ between two
eigenstates $|\phi _{k}{\rangle }$ and $|\phi _{j}{\rangle }$. Note that $%
X_{kj}=0$ if $|\phi _{k}{\rangle }$ and $|\phi _{j}{\rangle }$ are of the
same parity, providing the selection rule of the correlation measurement
induced eigenstate transition in the transition blockade $|\phi _{k}{\rangle
}{\nrightarrow }|\phi _{j}{\rangle }$.

{Note from Eq. (\ref{g2approx}) that $G_{2}(0)$ diverges as }$\Delta _{1,0}{%
\rightarrow }0$. Combining with small contributions of other unimportant
processes, the asymptotic trend of the energy level crossing between the ground
state and the first excited state lifts up the two-photon correlation
function, resulting in a peak structure in the $G_{2}(0)$ curve around $%
g/\omega _{0}{=}1.1$ in the dissipative AQRM, e.g., $r=0.5$.
For comparison, there are no any first-order QPT, i.e., those two lowest energy levels never cross, and are only closer with the
increasing coupling strength, thus the two-photon correlation
function varies smoothly and does not show a peak structure.
 Hence, this assures us that the giant photon-bunching signal is an emerging phenomenon when switching from the isotropic dissipative QRM to anisotropic case.


Except the photon-bunching peak in the middle of Fig.~\ref%
{fig1} (a), it is also intriguing to
find that the second critical coupling strength $g_{c}^{(2)}$ for the second
first-order QPT derived in Ref \ \cite{xychen2021pra} can also be roughly
captured by the photon-bunching feature with $r{\lesssim }0.25$, as shown in
Fig.~\ref{fig1}(b).
To describe the implication of the photon bunching peak on  the
first-order QPTs, we refine the detection signal of second first-order QPT by
modulating two bath temperatures in Fig.~\ref{fig4}. It is clearly shown
that with the decreasing temperature, the photon-bunching  range
monotonically shrink to the positions of the second and the third first-order
QPTs denoted by the dashed lines.

Then, we analytically connect photon-bunching feature with the first-order QPT
{of the } AQRM. As the first excited state approaches the ground state, $%
\Delta _{10}$ vanishes gradually. In the low temperature regime, two-photon
correlation function Eq.~(\ref{g2approx}) is approximated as
\begin{equation}
~G_{2}(0)\approx \frac{4(\Delta _{2,0}^{2}|X_{0,2}|^{2}+\Delta
_{2,1}^{2}|X_{1,2}|^{2})\Delta _{3,2}^{2}|X_{2,3}|^{2}P_{3}}{\Delta
_{1,0}^{4}|X_{0,1}|^{4}},
\end{equation}%
with $P_{1}{\approx }1/2$. As the qubit-photon coupling strength $g$
approaches $g_{c}^{(n)}$, one-photon term ${\langle }\hat{X}^{-}\hat{X}^{+}{%
\rangle }{\approx }\Delta _{1,0}^{2}|X_{0,1}|^{2}/2$ is strongly suppressed,
which directly leads to the pronounced photon-bunching feature.

We should note that from Ref.~\cite{xychen2021pra} that the numerical resolution of the
level crossings is very difficult, especially in the deep-strong coupling
regime, so it is challenging to identify  the multiple first-order QPTs using the energy spectra. However, the signal of the
two-photon correlation function, which is actually dependent on both the energy spectra and the eigenstates,  is pronounced and quite obvious. Since the
two-photon correlation function can be measured experimentally~\cite%
{afink2018np,dechang2014np,shc2020np}, we thus propose an alternative way to
detect the locations of the first-order QPTs in this work.

\section{\label{conclusion} Conclusion}

To summarize, we apply the quantum dressed master equation to investigate
first-order quantum phase transition via giant photon-bunching feature in the dissipative AQRM, which enables us to
properly treat the strong qubit-photon coupling.
{The two-photon correlation function at the steady state can be reasonably obtained, and it generally shows a giant photon-bunching peak at deep-strong
qubit-photon coupling, which is however
unavailable in the dissipative QRM.
Based on the numerical and analytical analysis, it is found that such photon-bunching peak is highly related with the emergence of the first-order quantum phase transition,
which is characterized as the level crossing of the ground state and
first-excited state. Since the two-photon correlation can be measured
experimentally, we suggest that measuring photon bunching signal would
characterize the first-order QPT of the qubit-photon systems. Finally, it
should be noted that the present results are irrelevant with system-bath
dissipation {strengths}, because weak interactions between the quantum
system components, i.e., qubit and photons, and the corresponding thermal
baths have been taken into account. }\\

{
We should admit that some deep-strong qubit-photon coupling strengths in this work, e.g., $g/\omega_0=2$, currently is experimentally unavailable based on the circuit quantum electrodynamics setups,
for the largest reported qubit-photon interaction strength is $g/\omega_0{\approx}1.34$~\cite{yoshihara2017np}.
However, with  the tremendous progress of the circuit quantum electrodynamics systems from strong~\cite{wallraff2004nat},
ultrastrong~\cite{diaz2010prl,niemczyk2010np}, to deep-strong couplings~\cite{yoshihara2017np},
we believe that the quantum technology of the superconducting circuits~\cite{blais2021rmp} may break through this bottleneck in near future.
As a consequence,
our theoretical exploration of interesting quantum optical phenomena could be observed,
which may provide insights for nonclassical photon statistics and the first-order quantum phase transition of the AQRM.
}
\section{Acknowledgements}

T. Y and Q.-H.C. are supported by the National Science
Foundation of China under Grant No. 11834005 and  the National
Key Research and Development Program of China under
Grant No. 2017YFA0303002. C.W. acknowledges the National
Natural Science Foundation of China under Grant No. 11704093 and the Opening
Project of Shanghai Key Laboratory of Special Artificial Microstructure
Materials and Technology.


\begin{thebibliography}{99}
\bibitem{scully1997book} M. O. Scully and M. S. Zubairy, \emph{Quantum Optics%
} (Cambridge University Press, Cambridge, 1997); P. Meystre, \emph{Quantum
Optics} (Springer, Cham, 2021).

\bibitem{haroche2006book} S. Haroche and J. M. Raimond, \emph{Exploring the
Quantum: Atoms, Cavities, and Photons} (Oxford University Press, Oxford,
2006).

\bibitem{wallquist2009pst} M. Wallquist, K. Hammerer, P. Rabl, M. Lukin, and
P. Zoller, Phys. Scr. \textbf{T137}, 014001 (2009).

\bibitem{kurizki2015pnas} G. Kurizki, P. Bertet, Y. Kubo, K. M{\o }lmer, D.
Petrosyan, P. Rabl, and J. Schmiedmayer, Proc. Natl. Acad. Sci. USA \textbf{%
112}, 3866 (2015).

\bibitem{diaz2019rmp} P. Forn-D\'{\i}az, L. Lamata, E. Rico, J. Kono, and E.
Solano, Rev. Mod. Phys. \textbf{91}, 025005 (2019).

\bibitem{kockum2019nrp} A. F. Kockum, A. Miranowicz, S. De Liberato, S.
Savasta, and F. Nori, Nat. Rev. Phys. \textbf{1}, 19 (2019).

\bibitem{blais2020np} A. Blais, S. M. Girvin, and W. D. Oliver, Nat.
Phys. \textbf{16}, 247 (2020).

\bibitem{clerk2020np} A. A. Clerk, K. W. Lehnert, P. Bertet, J. R. Petta,
and Y. Nakamura, Nat. Phys. \textbf{16}, 257 (2020).

\bibitem{hwang2015} M. J. Hwang, R. Puebla, and M. B. Plenio, Phys. Rev.
Lett. \textbf{115}, 180404 (2015).

\bibitem{Lin2017} M. X. Liu, S. Chesi, Z.-J. Ying, X. S. Chen, H.-G. Luo,
and H.-Q. Lin, Phys. Rev. Lett. \textbf{119}, 220601 (2017).

\bibitem{xychen2021pra} X. Y. Chen, L. W. Duan, D. Braak, and Q.-H. Chen,
Phys. Rev. A \textbf{103}, 043708 (2021).

\bibitem{Ying2021} Z. J. Ying, Adv. Quantum Technol. \textbf{4}, 2100088
(2021).

\bibitem{duan2021nc} M. L. Cai, Z. D. Liu, W. D. Zhao, Y. K. Wu, Q. X. Mei,
Y. Jiang, L. He, X. Zhang, Z. C. Zhou, and L. M. Duan, Nat. Commun. \textbf{%
12}, 1126 (2021).

\bibitem{rozani2018np} A. Ronzani, B. Karimi, J. Senior, Y. C. Chang, J. T.
Peltonen, C. D. Chen, and J. P. Pekola, Nat. Phys. \textbf{14}, 991 (2018).

\bibitem{senior2020cp} J. Senior, A. Gubaydullin, B. Karimi, J. T. Peltonen,
J. Ankerhold, and J. P. Pekola, Commun. Phys. \textbf{3}, 40 (2020).

\bibitem{pekola2021rmp} J. P. Pekola and B. Karimi, Rev. Mod. Phys. \textbf{%
93}, 041001 (2021).

\bibitem{cwang2021pra} C. Wang, H. Chen, and J. Q. Liao, Phys. Rev. A
\textbf{104}, 033701 (2021).

\bibitem{anappara2009prb} A. A. Anappara, S. De Liberato, A. Tredicucci, C.
Ciuti, G. Biasiol, L. Sorba, and F. Beltram, Phys. Rev. B \textbf{79},
201303(R) (2009).

\bibitem{niemczyk2010np} T. Niemczyk, F. Deppe, H. Huebl, E. P. Menzel, F.
Hocke, M. J. Schwarz, J. J. Garcia-Ripoll, D. Zueco, T. H\"{u}mmer, E.
Solano, A. Marx, and R. Gross, Nat. Phys. \textbf{6}, 772 (2010).

\bibitem{yoshihara2017np} F. Yoshihara, T. Fuse, S. Ashhab, K. Kakuyanagi,
S. Saito, and K. Semba, Nat. Phys. \textbf{13}, 44 (2017).

\bibitem{rabi_rabimodel} I. I. Rabi, Phys. Rev. \textbf{49}, 324 (1936).

\bibitem{braak2011prl} D. Braak, Phys. Rev. Lett. \textbf{107}, 100401
(2011).

\bibitem{chen2012pra} Q.-H. Chen, C. Wang, S. He, T. Liu, and K. L. Wang,
Phys. Rev. A \textbf{86}, 023822 (2012).

\bibitem{Zhong13} H. Zhong, Q. Xie, M. T. Batchelor, and C. Lee, J. Phys. A:
Math. Theor. \textbf{46}, 415302 (2013).

\bibitem{braak2016jpa} D. Braak, Q.-H. Chen, M. Batchelor, and E. Solano, J.
Phys. A: Math. Theor. \textbf{49}, 300301 (2016).

\bibitem{haroche2020np} S. Haroche, M. Brune, and J. M. Raimond, Nat. Phys.
\textbf{16}, 243 (2020).

\bibitem{jcm1963ieee} E. T. Jaynes and F. W. Cummings, Proc. IEEE \textbf{51}%
, 89 (1963).


\bibitem{ov2011prl} O. Viehmann, J. von Delft, and F. Marquardt, Phys.
Rev. Lett. 107, 113602 (2011).


\bibitem{baust2016prb} A. Baust, E. Hoffmann, M. Haeberlein, M. J. Schwarz,
P. Eder, J. Goetz, F. Wulschner, E. Xie, L. Zhong, F. Quijandr\'{\i}a, D.
Zueco, J.-J. Garc\'{\i}a Ripoll, L. Garca-\'{A}lvarez, G. Romero, E. Solano,
K. G. Fedorov, E. P. Menzel, F. Deppe, A. Marx, and R. Gross, Phys. Rev. B
\textbf{93}, 214501 (2016).

\bibitem{stockklauser2017prx} A. Stockklauser, P. Scarlino, J. V. Koski, S.
Gasparinetti, C. K. Andersen, C. Reichl, W. Wegscheider, T. Ihn, K. Ensslin,
and A. Wallraff, Phys. Rev. X \textbf{7}, 011030 (2017).

\bibitem{bayer2017nanolett} A. Bayer, M. Pozimski, S. Schambeck, D. Schuh,
R. Huber, D. Bougeard, and C. Lange, Nano Lett. \textbf{17}, 10, 6340 (2017).

\bibitem{jc2010prl} J. Casanova, G. Romero, I. Lizuain, J. J. Garc\'{\i}%
a-Ripoll, and E. Solano, Phys. Rev. Lett. \textbf{105}, 263603 (2010).

\bibitem{yyzhang2013cpl} Y.-Y. Zhang, Q.-H. Chen, S.-Y. Zhu, Chinese Physics
Letters \textbf{30}, 114203 (2013).

\bibitem{lg2016prl} L. Garziano, V. Macr\'{\i}, R. Stassi, O. Di Stefano, F.
Nori, and S. Savasta, Phys. Rev. Lett. \textbf{117}, 043601 (2016).

\bibitem{jqyou2017pra} Z. Chen, Y. Wang, T. Li, L. Tian, Y. Qiu, K. Inomata,
F. Yoshihara, S. Han, F. Nori, J. S. Tsai, and J. Q. You, Phys. Rev. A
\textbf{96}, 012325.

\bibitem{Ye13} Y. X. Yu, J. Ye, and W. M. Liu, Sci. Rep. 3, 3476 (2013).

\bibitem{qtxie2014prx} Q. T. Xie, S. Cui, J. P. Cao, L. Amico, and H. Fan,
Phys. Rev. X \textbf{4}, 021046 (2014).

\bibitem{ab2014prl} A. Baksic and C. Ciuti, Phys. Rev. Lett. \textbf{112},
173601 (2014).

\bibitem{wjy2017pra} W.-J. Yang and X.-B. Wang, Phys. Rev. A \textbf{95},
043823 (2017).

\bibitem{gcwang2019scirep} G. C. Wang, R. Q. Xiao, H. Z. Shen, C. F. Sun,
and K. Xue, Sci. Rep. \textbf{9}, 4569 (2019).

\bibitem{erlingsson2010pra} S. I. Erlingsson, J. C. Egues, and D. Loss, Phys.
Rev. B \textbf{82}, 155456 (2010).

\bibitem{skogvoll2021arxiv} I. C. Skogvoll, J. Lidal, J. Danon, and A. Kamra,
Phys. Rev. Applied \textbf{16}, 064008 (2021).

\bibitem{uweiss2012book} U. Weiss, \emph{Quantum Dissipative Dynamics}
(World Scientific, Singapore, 2012).

\bibitem{schmidt2010prb} S. Schmidt, D. Gerace, A. A. Houck, G. Blatter, and
H. E. T\"{u}reci, Phys. Rev. B \textbf{82}, 100507(R) (2010).

\bibitem{hwang2016prl} M. J. Hwang, M. S. Kim, and M.-S. Choi, Phys. Rev.
Lett. \textbf{116}, 153601 (2016).

\bibitem{hwang2018pra} M. J. Hwang, P. Rabl, and M. B. Plenio, Phys. Rev. A
\textbf{97}, 013825 (2018).

\bibitem{hjc2015prx} H. J. Carmichael, Phys. Rev. X \textbf{5}, 031028
(2015).

\bibitem{jfink2017prx} J. M. Fink, A. Dombi, A. Vukics, A. Wallraff, and P.
Domokos, Phys. Rev. X \textbf{7}, 011012 (2017).

\bibitem{hjc1999book} H. J. Carmichael, \emph{Statistical Methods in Quantum
Optics 1, Master Equations and Fokker Planck Equations} (Springer, 1999).

\bibitem{rjg1963pr} R. J. Glauber, Phys. Rev. \textbf{130}, 2529 (1963).

\bibitem{blais_dME} F. Beaudoin, J. M. Gambetta, and A. Blais, Phys. Rev. A
\textbf{84}, 043832 (2011).

\bibitem{alb2017pra} A. Le Boit\'{e}, M. J. Hwang, and M. B. Plenio, Phys.
Rev. A \textbf{95} 023829 (2017).

\bibitem{as2018pra} A. Settineri, V. Macr\'{\i}, A. Ridolfo, O. Di Stefano,
A. F. Kockum, F. Nori, and S. Savasta, Phys. Rev. A \textbf{98}, 053834
(2018).

\bibitem{alb2020aqt} A. Le Boit{\'{e}}, Adv. Quantum Technol. \textbf{3},
1900140 (2020); M.-S. Choi, \textsl{ibid}. \textbf{3}, 2000085 (2020).

\bibitem{ridolfo_photonblockade} A. Ridolfo, M. Lieb, S. Savasta, and M. J.
Hartmann Phys. Rev. Lett. \textbf{109} 193602 (2012).

\bibitem{ar2013prl} A. Ridolfo, S. Savasta, and M. J. Hartmann, Phys. Rev.
Lett. \textbf{110}, 163601 (2013).

\bibitem{alb2016pra} A. Le Boit\'{e}, M. J. Hwang, H. C. Nha, and M. B.
Plenio, Phys. Rev. A \textbf{94}, 033827 (2016).

\bibitem{lg2017acsp} L. Garziano, A. Ridolfo, S. De Liberato, and S.
Savasta, ACS Photonics \textbf{4}, 2345 (2017).



%



%
\bibitem{qs2019prl} Q. Schaeverbeke, R. Avriller, T. Frederiksen, and F.
Pistolesi, Phys. Rev. Lett. \textbf{123}, 246601 (2019).

\bibitem{shc2020np} S. H. Cantu, A. V. Venkatramani, W. C. Xu, L. Zhou, B.
Jelenkovi\'{c}, M. D. Lukin, and V. Vuleti\'{c}, Nat. Phys. \textbf{16}, 921
(2020).

\bibitem{asp2020np} A. S. Prasad, J. Hinney, S. Mahmoodian, K. Hammerer, S.
Rind, P. Schneeweiss, A. S. S{\o }rensen, J. Volz, and A. Rauschenbeutel,
Nat. Photon. \textbf{14}, 719 (2020).

\bibitem{qbin2020prl} Q. Bin, X.-Y. L\"{u}, F. P. Laussy, F. Nori, and Y.
Wu, Phys. Rev. Lett. \textbf{124}, 053601 (2020).

\bibitem{afink2018np} T. Fink, A. Schade, S. H\"{o}fling, C. Schneider, and
A. Imamoglu, Nat. Phys. \textbf{14}, 365 (2018).

\bibitem{dechang2014np} D. E. Chang, V. Vuleti\'{c}, and M. D. Lukin, Nat.
Photon. \textbf{8}, 685 (2014).




{
\bibitem{wallraff2004nat} A. Wallraff, D. I. Schuster, A. Blais, L. Frunzio, R.-S. Huang,
J. Majer, S. Kumar, S. M. Girvin, and R. J. Schoelkopf, Nature, \textbf{431}, 162 (2004).
\bibitem{diaz2010prl} P. Forn-D\'{i}az, J. Lisenfeld, D. Marcos, J. J. Garc\'{i}a-Ripoll,
E. Solano, C. J. P. M. Harmans, and J. E. Mooij, Phys. Rev. Lett. \textbf{105}, 237001 (2010).
\bibitem{blais2021rmp} A. Blais, A. L. Grimsmo, S. Girvin, and A. Wallraff, Rev. Mod. Phys. \textbf{93}, 025005 (2021).}


\end{thebibliography}
\end{document}